\documentclass[runningheads]{llncs}
\usepackage{graphicx}
\usepackage{cite}
\usepackage{comment} 
\usepackage{amsmath,amssymb,amsfonts}
\usepackage{algorithmic}
\usepackage{graphicx}
\usepackage{textcomp}
\usepackage{xcolor}
\usepackage{subcaption} 
\usepackage{siunitx}

\begin{document}
\title{Experimental Characterization of Biological Tissue Dielectric Properties through THz Time-Domain Spectroscopy}
\titlerunning{THz-TDS for Biological Tissue Characterization}

\author{Elisabetta Marini\inst{1}\orcidID{0009-0003-1081-7815} \and
Silvia Mura\inst{1}\orcidID{0000-0002-0207-5730} \and
Marco Hernandez\inst{2} \and Matti H\"am\"al\"ainen\inst{2}\orcidID{0000-0002-6115-5255} \and Maurizio Magarini\inst{1}\orcidID{0000-0001-9288-0452}}
\authorrunning{E. Marini et al.}
%
\institute{$^1$ Dept. of Electronics, Information and Bioengineering, Politecnico di Milano, Italy\\
\email{\{elisabetta.marini,silvia.mura,maurizio.magarini\}@polimi.it}\\
$^2$ University of Oulu, Oulu, Finland\\
\email{\{marco.hernandez,matti.hamalainen\}@oulu.fi}}
\maketitle              
\begin{abstract}
Terahertz (THz) radiation provides a non-ionizing, highly sensitive probe of the dielectric properties of biological tissues. In this study, we present a comprehensive experimental characterization of dielectric properties using pork skin tissue, a widely used surrogate for human tissue, as a biological sample. Measurements are conducted employing THz time-domain spectroscopy in the 0.1–11~THz frequency range with photoconductive antennas for both signal generation and detection. Frequency-dependent refractive indices, absorption, and complex permittivity are extracted from transmitted time-domain signals. Our results confirm strong absorption and low transmittance at low THz frequencies due to water content, while highlighting frequency-dependent dispersion and narrowband transmission features at higher frequencies. This work provides one of the first extended-frequency datasets of biological tissue dielectric properties, supporting realistic channel modeling for the design and development of intra-body nanosensor networks in the THz band. 


\end{abstract}

\section{Introduction}

The integration of nanoscale devices within the human body has the potential to radically transform modern medicine. By enabling continuous physiological monitoring, targeted therapies, and seamless interaction with biological systems, these nanosystems pave the way for personalized healthcare, early disease detection, and advanced biomedical interfaces~\cite{GHORBIAN2025100740}. The vision of intra-body wireless nanosensor networks (WNSNs) encompasses numerous devices capable of sensing, processing, and communicating biomedical information in real time, thus supporting applications as diverse as precision drug delivery, neural interfacing for brain–machine interfaces (BMIs), and monitoring of chronic diseases~\cite{naranjo2018past,vizziello2023intra}.

A key enabler of this vision is intra-body communication (IBC), defined as the reliable and efficient exchange of information among nanoscale devices and between nanosensors and external gateways. Communication requirements are stringent: high data rates are needed for neural data acquisition, ultra-low latency is essential for closed-loop biomedical control, and power consumption must stay very low for safe, sustainable operation. These challenges have driven research into novel communication paradigms for the intra-body medium~\cite{naranjo2018past}.

Several technological approaches have been proposed to realize intra-body communication (IBC). Molecular communication (MC), exploiting controlled release and detection of molecules, offers excellent biocompatibility and minimal energy consumption~\cite{vizziello2023intra,anjum2019molecular}. However, the stochastic nature of molecular diffusion causes high latency and limited data rates, restricting its use to delay-tolerant applications such as long-term monitoring or drug delivery. Ultrasonic communication provides improved propagation in tissues and potentially higher data rates. Yet, challenges such as device miniaturization, high power requirements, and limited integration with nanoscale architectures remain significant~\cite{naranjo2018past,vizziello2023intra}.  

Electromagnetic (EM) communication is another promising solution, benefiting from decades of wireless technology progress. Different frequency bands have contrasting trade-offs: low frequencies (kHz–MHz) penetrate deeply but offer limited data rates; RF and microwave bands provide higher bandwidths, but tissue absorption reduces range; communication at millimeter waves offers wide bandwidths, yet strong attenuation in water-rich tissues severely limits practical range~\cite{vizziello2023intra}. These constraints have motivated the search for even higher-frequency bands that could combine wide bandwidth with sufficient tissue penetration.  

In this context, the terahertz (THz) band (0.1–10~THz) emerges as a particularly compelling candidate for IBC. THz radiation provides ultra-wide bandwidths for very high data rates and low latency, while remaining non-ionizing and biologically safer than X-rays or other high-energy radiation. Recent advances demonstrate the feasibility of nanoscale transceivers resonating at THz frequencies, bridging the gap between miniaturization and communication performance~\cite{hanson2008fundamental,jornet2013fundamentals,akyildiz2010electromagnetic}. These developments position the THz band as a central enabler for intra-body wireless nanosensor networks (WNSNs).

Despite these advantages, challenges remain. Biological tissues, rich in water, strongly absorb THz radiation, causing significant signal attenuation~\cite{guo2016channel}. Tissue heterogeneity leads to complex scattering and dispersion~\cite{costa2018in}. Careful power management, thermal control, and optimized device placement are needed for safe and reliable operation. Accurate channel modeling that includes absorption, scattering, and dispersion is therefore essential~\cite{jornet2011channel,guo2016channel,jornet2013fundamentals}. Notably, studies on THz propagation in tissues often focus on narrower frequency ranges, such as 0.1–3.5 THz~\cite{Peretti2019} or higher bands up to 20 THz~\cite{arxiv2020THz}, using techniques like frequency-domain spectroscopy or THz imaging.

Considering these limitations, it is clear that more comprehensive experimental data across the full THz range are needed for intra-body communication. To address this, the present work investigates THz propagation in porcine skin, a widely used surrogate for human tissue because of its structural and dielectric similarities. By employing THz time-domain spectroscopy (TDS) with photoconductive antenna (PCA) instrumentation, key dielectric parameters, including refractive index, absorption coefficient, and dielectric constant, are extracted across 0.1 to 10~THz. These parameters provide realistic inputs for channel modeling and highlight trade offs between achievable bandwidth and penetration depth in intra-body environments.

The paper is organized as follows. Section~\ref{sec:systemModel} presents the system model. Section~\ref{sec:terahz_rad} covers THz propagation in tissues, Sec.~\ref{sec:experimentalcampaign} describes the experimental setup, and Sec.~\ref{sec:numericalres} reports the results. Finally, Sec.~\ref{sec:conclusion} concludes the paper.
\begin{figure}[!b]
    \centering
    \includegraphics[width=.93\linewidth]{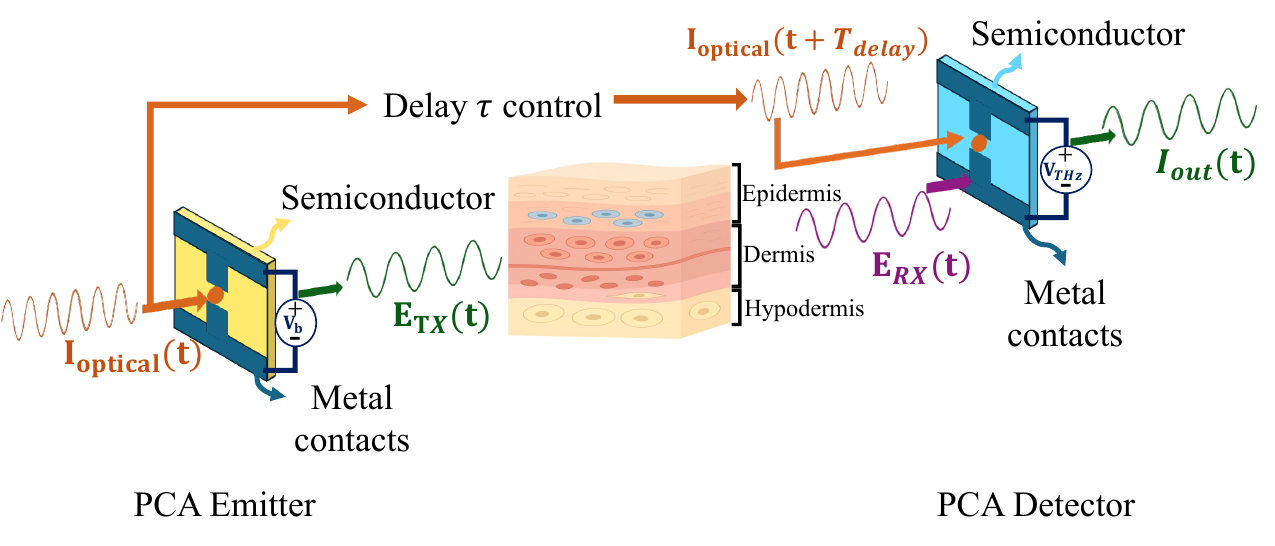}
    \vspace*{-.3cm}
    \caption{Figurative diagram of the THz time-domain spectroscopy system for tissue analysis. The THz pulse, generated and detected using PCAs, interacts with the tissue, and the response is processed to reconstruct the waveform.}
    \label{fig: sys_mod}
\end{figure}

\section{System Model}\label{sec:systemModel}

The interactions of THz radiation with cells and molecules in biological tissues are studied using TDS with PCAs for broadband generation and detection. Figure~\ref{fig: sys_mod} shows the system workflow, providing a schematic of the experimental THz time-domain chain for biological tissue applications.

\subsection{Photoconductive antenna (PCA)}

A PCA is typically fabricated on a semiconductor substrate, such as GaAs or InGaAs, with two metallic electrodes separated by a narrow gap of width $d$. The semiconductor has a valence band (VB) and a conduction band (CB), separated by an energy gap $E_G$. At equilibrium, electrons populate the VB, ensuring charge neutrality. When illuminated by a femtosecond laser pulse with photon energy exceeding $E_G$, electrons are excited into the CB, generating electron–hole pairs. Under an external bias, these carriers accelerate, producing a transient photocurrent that generates THz radiation~\cite{THz_MC,THz_model}. The emitted THz field, $E_{TX}(t)$, can be modeled as the emitter response, $h_{emitter}(t)$, to the optical excitation combined with the bias voltage. In detection, a delay $\tau$ is introduced to the optical path, allowing the transmitted THz pulse, $E_{RX}(t)$, to interact with the detector PCA, described by $h_{detector}(t)$, enabling time-domain sampling of the waveform.  

Pulse characteristics depend on semiconductor parameters such as carrier lifetime, mobility, and saturation velocity, as well as optical pulse duration and bias field. Electrode geometry (e.g., dipole, bow-tie, or interdigitated structures) further influences the radiation pattern and efficiency, providing spectral tunability for specific applications.

\subsection{THz signal generation}
The PCA converts ultra-fast optical pulses into electromagnetic radiation in the THz domain, and its behavior is modeled by the temporal response $h_{emitter}(t)$. When the PCA operates as an emitter, a static bias voltage, $V_b \,= 10\, V$, is applied across the electrodes, establishing a static electric field in the gap region $E_{bias}$. A laser pulse of duration $\tau_{pulse} $$\,=\,$$ 10\,$--$\,100\,$fs, typically delivered by a pulsed $Ti:sapphire$ laser at a wavelength of $\lambda$$\,=\,$$800 \,n$m, irradiates a GaAs substrate of thickness $L$$\,=\,$$0.1\,$mm across a gap of $d$$\,=\,$$5\,\mu$m between the two metal contacts. The optical excitation has an average power $P_{avg}$ of a few hundred mW, and a repetition frequency $f_r$$\,=\,$$80\,$MHz~\cite{THz_app}.

The photons of the focused femtosecond laser  carry an energy $E_{laser}$$\,=\,$$hf$, where $f$$\,=\,$$\frac{c}{\lambda}$ is the optical frequency determined by the speed of light $c$ and the wavelength $\lambda$. This photon energy exceeds the semiconductor band gap $E_G$, thereby generating electron–hole pairs. These carriers are accelerated by the static field, producing a transient photocurrent $I_p(t)$. The carrier density $n(t)$ follows dynamics that can be modeled as~\cite{Wilson2023,Qin2021,Piao1999}
\begin{equation}
\frac{dn(t)}{dt} = G(t)-\frac{n(t)}{\tau_c},
\end{equation}
where $G(t)$ is the carrier generation rate and $\tau_c$ the carrier recombination lifetime. The average carrier density $\overline{n}$ can be estimated from the number of photons absorbed in the active volume of the device. Specifically, it can be expressed as
\begin{equation}
\overline{n} \simeq  \frac{N_p}{V_A},
\end{equation}
where $V_A$ is the active volume and $N_p$ the number of photons per pulse, reported in~\cite{Wahlstrand2018} as
\begin{equation}
N_p = \frac{E_\mathrm{pulse}}{h\nu} = \frac{P_\mathrm{avg}/f_r}{hc/\lambda} \simeq 10^{8},
\end{equation}
with $h$ denoting Planck's constant. The induced photocurrent can then be written as~\cite{Wahlstrand2018,Castro-Camus2005}
\begin{equation}
I_p(t) = A\, E_{bias}\,q\,\mu\,n(t)\simeq 10^{-2}\,A,
\end{equation}
where $A$ is the beam-illuminated area, $q$ the elementary charge, and $\mu$ the mobility.

The THz radiation field emitted in the far field is proportional to the temporal derivative of the photocurrent~\cite{Piao1999,Castro-Camus2005,Wahlstrand2018}:
\begin{equation}
E_{\mathrm{TX}}(t)\propto \frac{d I_p(t)}{dt}.
\end{equation}
Considering the femtosecond pulse duration, and taking into account the geometry of the PCA and propagation to the far-field, the corresponding radiated field is on the order of
\begin{equation}
    E_{\mathrm{TX}}(t) \simeq 10^{4}-10^{6}\, V/m.
    \label{eq:ETX}
\end{equation}

These relations link the optical excitation, material properties, and antenna geometry to the amplitude and temporal profile of the emitted THz waveform.

\subsection{THz signal detection}

When a PCA operates as a detector of THz radiation, its response, $h_{detector}(t)$, converts the incident THz pulse from the sample, $E_{RX}(t)$, into a photocurrent $I_\mathrm{det}(t)$ that carries information about the sample properties. In a PCA detector, a delayed portion of the same femtosecond laser pulse generates electron–hole pairs in the semiconductor. The incident THz field $E_\mathrm{RX}(t)$ accelerates these carriers, producing a time-dependent photocurrent $I_\mathrm{det}(t)$ proportional to the THz signal. This photocurrent can be expressed as~\cite{Burford2017}
\begin{equation}
I_{det}(t) = g_d(t) \cdot V_{\mathrm{THz}}(t),
\end{equation}
where $g_d(t)$ is the time-dependent photoconductance determined by the optical gating pulse and the carrier lifetime, and $V_{\mathrm{THz}}(t)$ is the voltage induced across the antenna gap by the incident THz radiation. The induced voltage $V_{\mathrm{THz}}(t)$ is proportional to the incident THz field, with the proportionality set by the effective antenna length $L_{\mathrm{eff}}$~\cite{Burford2017}, yielding
\begin{equation}
V_{\mathrm{THz}}(t) = L_{\mathrm{eff}} \cdot E_{\mathrm{RX}}(t).
\end{equation}

Therefore, the output photocurrent as a function of the optical delay $T_{delay}$, which represents the relative temporal delay between the pump and probe optical pulses, can be written as~\cite{Piao1999,Castro-Camus2005,Wahlstrand2018}
\begin{equation}
I_{\text{out}}(t) \propto E_{\text{RX}}(t+T_{delay}).
\end{equation}
The delay $T_{\text{delay}}$ is a controlled, constant parameter for each acquisition point, independent of the detection channel. This preserves both the magnitude and phase of the THz radiation transmitted or reflected by the sample, enabling extraction of its complex transmission (or reflection) coefficient. The detected current is then amplified and processed by the readout electronics. By scanning the optical delay, the complete temporal waveform of $E_{\text{RX}}(t)$ is reconstructed. A Fourier transform provides the spectral amplitude and phase, from which material parameters such as refractive index, absorption coefficient, and dielectric constant are derived to characterize the tissue response to THz radiation.

\subsection{System operation for intra-body analysis}
The intra-body PCA-based THz system operates by sequentially generating, propagating, and detecting pulses to analyze biological tissues. The emitted pulse travels through tissue, experiencing absorption, dispersion, and scattering due to the tissue's dielectric properties. Tissue effectively modulates both the amplitude and phase of the transmitted or reflected waveform. The recorded signal encodes absorption and phase delay induced by the tissue, enabling quantitative and qualitative analysis of intra-body propagation. Temporal propagation within the biological tissue can be modeled as
\begin{equation}
E_{\mathrm{RX}}(t) = E_{\mathrm{TX}}(t) *  h_{\mathrm{tissue}}(t) ,
\end{equation}
where $*$ is the convolution operation and  $h_{\mathrm{tissue}}(t)$ is the time-dependent tissue channel response accounting for absorption, dispersion, and scattering, quantified with the tissue's complex permittivity $\epsilon$.
\begin{figure}[!b]
    \centering
    \includegraphics[width=.95\linewidth]{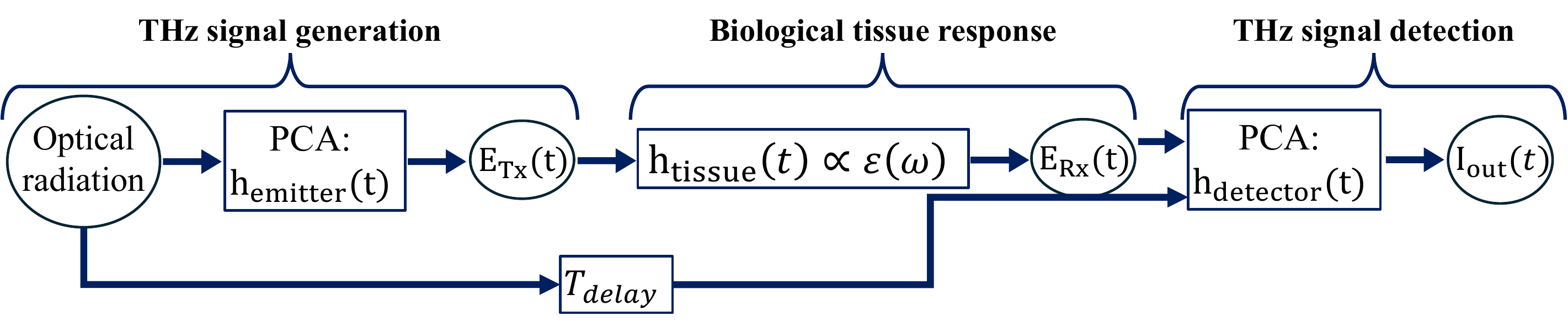}
    \vspace*{-.3cm}
    \caption{Stages of the proposed framework. The experiment comprises three main steps: signal generation, tissue interaction, and signal detection.}
    \label{modello}
\end{figure}

Figure~\ref{modello} details the three main stages of the considered system and the corresponding operations for each block:
\begin{enumerate}
    \item \textbf{THz signal generation}: The THz radiation is generated by the PCA, whose response to the incoming optical radiation is described by the impulse response $h_{emitter}(t)$.
    \item \textbf{Biological tissue response}: The tissue interacts with the incident electromagnetic field $E_{TX}(t)$, resulting in $E_{RX}(t)$. This interaction can be described by the tissue channel response $h_{tissue(t)}$, which depends on the tissue’s absorption coefficient and dielectric properties~\cite{Pickwell2006};
    \item \textbf{THz signal detection}: as the resulting radiation $E_{RX}(t)$ is carrying the information on tissue properties, it is converted in an electric measurable quantity from which it is possible to extract the dielectric parameters.
\end{enumerate}

\section{Terahertz Propagation in Biological Tissues}\label{sec:terahz_rad}

The propagation of THz radiation in biological tissues is determined by their complex dielectric properties, which govern energy storage, dissipation, and attenuation. The THz range (0.1--10~THz) is highly sensitive to water content and molecular vibrations, making it an effective probe for soft tissues, hydration, and cellular organization~\cite{Pickwell2006}. At the molecular level, biological function depends on weak interactions, such as hydrogen bonds and Van der Waals forces, which shape biomolecular assembly and tissue structure~\cite{Wang2019NanoBio,Cherkasova2021}. The energy of THz photons (0.4--40~meV) overlaps with rotational, vibrational, and collective molecular modes~\cite{Cherkasova2021}, revealing dynamics often inaccessible to higher-energy techniques like infrared spectroscopy. Understanding the dielectric response in this range is crucial, linking molecular dynamics to macroscopic electromagnetic behavior and defining absorption, dispersion, and propagation.

When radiation propagates through a material of thickness $d$, the resulting outgoing intensity follows Beer--Lambert's law~\cite{Shi2022}:
\begin{equation}
|E_{RX} (\omega)| = |E_{TX}(\omega)| e^{-\alpha(\omega) d},
\end{equation}
where $\alpha(\omega)$ is the absorption coefficient. Experimental characterization is commonly performed using THz time-domain spectroscopy (THz-TDS), which enables direct measurement of the complex transmission coefficient~\cite{Peretti2019}:
\begin{equation}
T(\omega) = \frac{E_{\text{RX}}(\omega)}{E_{\text{TX}}(\omega)}.
\end{equation}

From the transmittance, tissue parameters can be extracted, as they depend on the magnitude of $T(\omega)$ and on the phase difference $\phi(\omega) = \arg \left( \frac{E_{RX}(\omega)}{E_{TX}(\omega)} \right)$~\cite{Peretti2019,Calvo2021,Jepsen2019}. The absorption coefficient $\alpha(\omega)$ and the refractive index $n(\omega)$ are obtained as
\begin{equation}
\alpha(\omega) \approx -\frac{1}{d}\ln \left( \frac{|E_{RX}(\omega)|}{|E_{TX}(\omega)|} \right), \quad
n(\omega) = 1 + \frac{c\, \phi(\omega)}{\omega d}.
\end{equation}
The extinction coefficient $k(\omega)$, which quantifies the material's electromagnetic losses, is expressed as
\begin{equation}
k(\omega) = \frac{c\, \alpha(\omega)}{2 \, \omega}.
\end{equation}

Using $n(\omega)$ and $k(\omega)$, the complex permittivity is obtained through
\begin{equation}
\tilde{\varepsilon}_r(\omega) = \left[ n(\omega) - i k(\omega) \right]^2 = \varepsilon'_r(\omega) - i \varepsilon''_r(\omega),
\end{equation}
where $\varepsilon'_r(\omega)$ represents the real part, the dielectric constant, representing the material’s ability to store energy, and and $\varepsilon''_r(\omega)$ represents the imaginary component, the dielectric loss, quantifying energy dissipation through absorption and relaxation mechanisms. This is possible thanks to the direct dependence of the absorption coefficient $\alpha(\omega)$ and the refractive index $n(\omega)$ on the amount of the transmitted radiation with respect to the received amount.

The dielectric response of tissues is strongly influenced by water content, temperature, and molecular composition. For water-rich soft tissues, empirical relationships link static permittivity $\varepsilon_s$, high-frequency permittivity $\varepsilon_\infty$, and relaxation time $\tau$ to the water mass fraction $W_c$:
\begin{equation}
\varepsilon_s = a_1 + b_1 W_c, \quad
\varepsilon_\infty = a_2 + b_2 W_c, \quad
\tau = a_3 + b_3 W_c,
\end{equation}
where \( a_1, a_2, a_3 \) represent the baseline values for a dry or low-water-content medium, while \( b_1, b_2, b_3 \) describe the rate of variation of each parameter with the water mass fraction \( W_c \).
Finally, the frequency-dependent dielectric function can be modeled using a generalized Debye relaxation to account for both bulk and bound water contributions \cite{jornet}:
\begin{equation}
\tilde{\varepsilon}_r(\omega) = \varepsilon_\infty + \sum_{j=1}^{N} \frac{\Delta \varepsilon_j}{1 + i \omega \tau_j},
\end{equation}
where $\Delta \varepsilon_j = \varepsilon_{sj} - \varepsilon_{\infty j}$ and $\tau_j$ represent the relaxation strength and relaxation time of the $j$th process, respectively.

Accurate characterization of the THz channel in biological tissues, through the determination of the real and imaginary parts of the relative permittivity, $\varepsilon'_r(\omega)$ and $\varepsilon''_r(\omega)$, as well as the refractive index $n(\omega)$, extinction coefficient $k(\omega)$, and absorption coefficient $\alpha(\omega)$, is essential. It enables reliable interpretation of THz imaging contrast, quantitative evaluation of tissue hydration, and understanding of molecular interactions involving water and proteins. Such knowledge is crucial for designing biocompatible THz systems with minimal energy deposition and signal loss in biological media. From these considerations, it emerges that these systems are particularly well suited for investigating multilayered biological tissues, as the analysis of both transmission and reflection allows the extraction of relevant parameters, including microstructural composition. This capability makes PCA-based THz spectroscopy a powerful, non-invasive tool for dermatological research.

\section{Experimental Framework}\label{sec:experimentalcampaign}
This study presents a detailed investigation of the electromagnetic properties of pork skin using the TeraPulse 4000 THz-TDS system. The measurement campaign aimed to characterize the tissue’s dielectric properties, refractive index, and absorption behavior in the 0.01--11~THz range, with the emitted pulse centered at 0.5~THz. The TeraPulse 4000 operated in transmission mode, generating femtosecond laser pulses centered at 800~nm with a spectral resolution of 1~GHz. The average optical power was 800~mW at an 80~MHz repetition rate. The THz beam was focused to a 3~mm spot with a pulse duration of 100~fs. Biological samples were positioned between the emitter and detector to maintain a constant optical path. The main system specifications are summarized in Table~\ref{tab:terapulse_specs}.
\begin{table}[!t]
\centering
\caption{TeraPulse 4000 system specifications}
\label{tab:terapulse_specs}
\begin{tabular}{|l|c|}
\hline
Parameter & Value \\
\hline
\hline
Frequency range & 0.01--11 THz \\
\hline
Spectral resolution & 1 GHz \\
\hline
Dynamic range & $>$90 dB \\
\hline
Laser wavelength & 800 nm \\
\hline
Pulse duration & 100 fs \\
\hline
Repetition rate & 80 MHz \\
\hline
Average power & 800 mW \\
\hline
Beam diameter & 3 mm \\
\hline
\end{tabular}
\end{table}
Pork skin can be selected as a representative model for human skin due to its comparable anatomical structure and physiological properties, making it a suitable proxy for biomedical investigations in imaging, spectroscopy, and therapeutic applications. The observed THz pulse reflects the dispersive behavior associated with the frequency-dependent dielectric response of the skin. Comparison between the measured parameters of pork skin and reported values for human skin~\cite{Sullivan2001} shows close agreement, confirming its relevance as a surrogate tissue in THz studies. As summarized in Table~\ref{tab:comparison_human}, the quantitative similarity between pork and human skin parameters, with only minor deviations due to hydration and thickness differences, further supports the use of pork skin as a reliable biological analog in THz dielectric characterization.

\begin{figure}[!t]
    \centering
    \begin{subfigure}{0.32\linewidth}
        \centering
        \includegraphics[width=.94\linewidth]{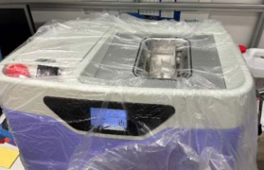}
        \caption{}
    \end{subfigure}
    \hfill
    \begin{subfigure}{0.35\linewidth}
        \centering
        \includegraphics[width=.94\linewidth]{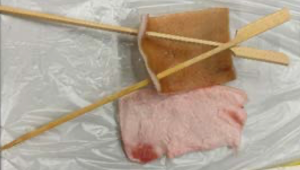}
        \caption{}
    \end{subfigure}
    \hfill
    \begin{subfigure}{0.28\linewidth}
        \centering
        \includegraphics[width=.94\linewidth]{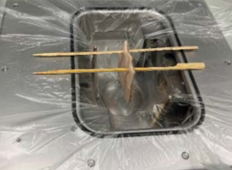}
        \caption{}
    \end{subfigure}
    \vspace*{-.3cm}
    \caption{Experiment setup overview. a) The instrumentation was covered with plastic to avoid contamination. b) Skin samples were hooked up to wooden sticks. c) Skin samples in the instrument chamber to perform the experiments,}
    \label{fig:confronto}
\end{figure}

In our experiments, fresh pork skin samples were collected post-mortem and stored at 4$^\circ$C to preserve their native properties. Preparation involved removing the subcutaneous fat layer and trimming the tissue to $5$$\,\times\,$$7\,$cm$^2$ with a thickness of about 2–3~mm, depending on fat nonuniformity. Measurements were performed immediately to prevent dehydration. The samples were mounted on wooden sticks inside the chamber, which was covered with a thin plastic layer to prevent contamination, as shown in Fig.~\ref{fig:confronto}. 
\begin{table}[!b]
\centering
\caption{Comparison between measured parameters of pork skin and literature values for human skin at [0-2]THz frequency range}
\label{tab:comparison_human}
\begin{tabular}{|c|c|c|}
\hline
Parameter & Pork skin~\cite{He2006,Helminiak2023} & Human skin~\cite{Peralta2019,Sullivan2001,Piro2016} \\
\hline
Absorption & 30--200 cm$^{-1}$ & 70--300 \\
Refractive Index & 1.7--2.4 & 1.5--2.6 \\
Dielectric Constant (Real part) & 2--3 & 2--4 \\
\hline
\end{tabular}
\end{table}
Both reference and sample time-domain signals were then recorded over a 30~ps window with 0.1~ps resolution. Averaging over 1000 scans was performed to improve the signal-to-noise ratio. The complex transmission coefficient $\hat{T}(\omega)$ was obtained by Fourier transforming the time-domain signals, enabling extraction of the amplitude and phase of the transmitted pulse. From these, the frequency-dependent complex refractive index, $\hat{n}(\omega) - i \hat{\kappa}(\omega)$, was retrieved. The real part $\hat{n}(\omega)$ described the phase velocity, while the imaginary part $\hat{\kappa}(\omega)$ represented absorption losses through $\hat{\alpha}(\omega) = 2\omega \hat{\kappa}(\omega)/c$.

The reported spectra may include minor uncertainties affecting the absolute magnitudes shown. These can arise from internal scaling or normalization by the instrument, as well as practical factors such as sample thickness, alignment, or phase processing. Small variations may occur when comparing data across studies due to differences in calibration, data processing, or experimental conditions (e.g., temperature, hydration, or humidity). Nevertheless, the spectra provide a consistent and reliable reference for analyzing the dielectric properties of hydrated biological tissues. The following section discusses the measurements in terms of transmittance, absorption coefficient, and dielectric constant.
\begin{figure}[!b]
    \centering
    \includegraphics[width=0.45\linewidth]{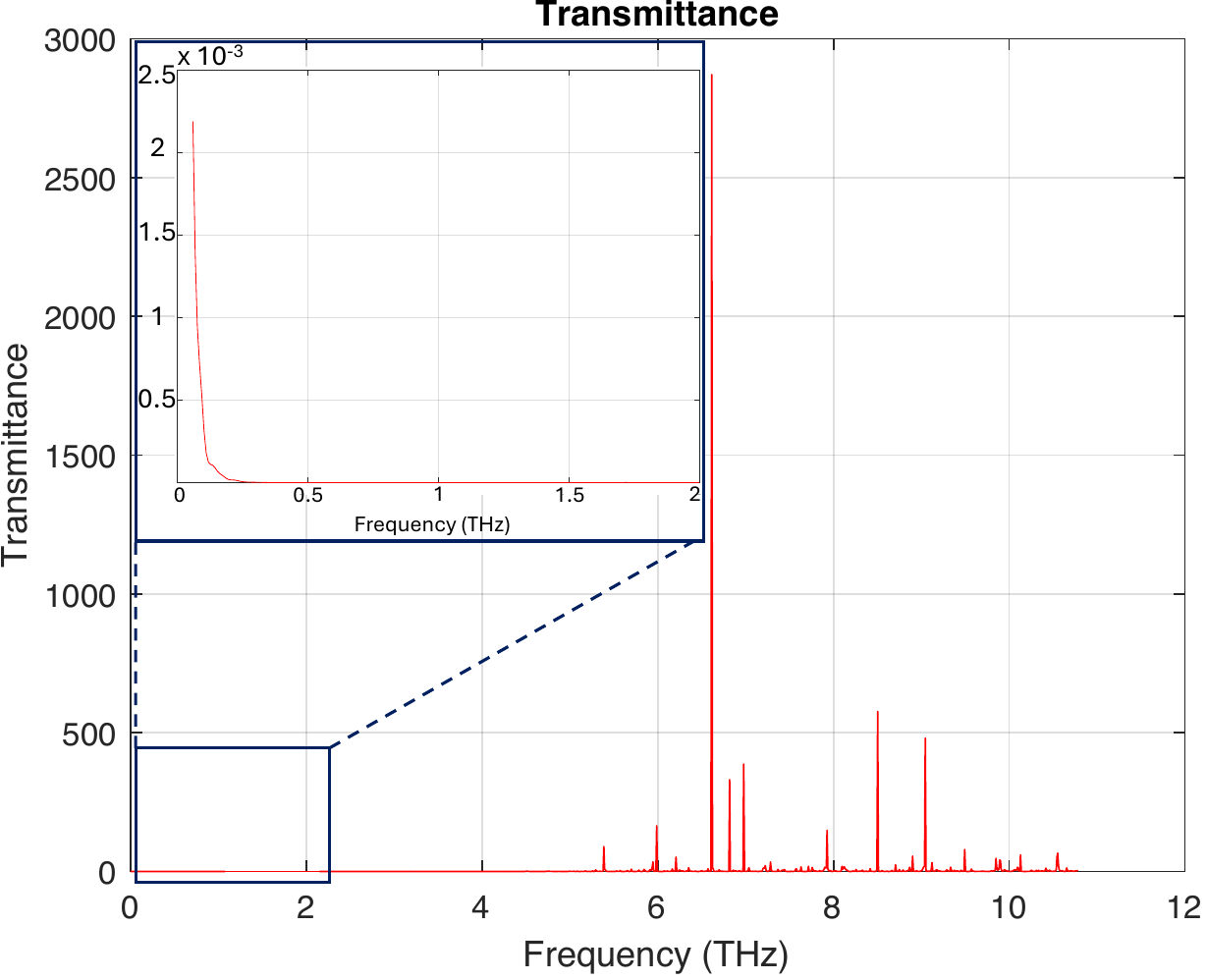}
    \vspace*{-.3cm}
    \caption{Measured transmittance in the [0--12]~THz range. A zoomed figure shows the behavior near the incident radiation frequency.}
    \label{trasm}
\end{figure}

\section{Numerical results}\label{sec:numericalres}
\begin{figure}[!b]
    \centering
        \begin{subfigure}{0.47\linewidth}
        \centering
        \includegraphics[width=.93\linewidth]{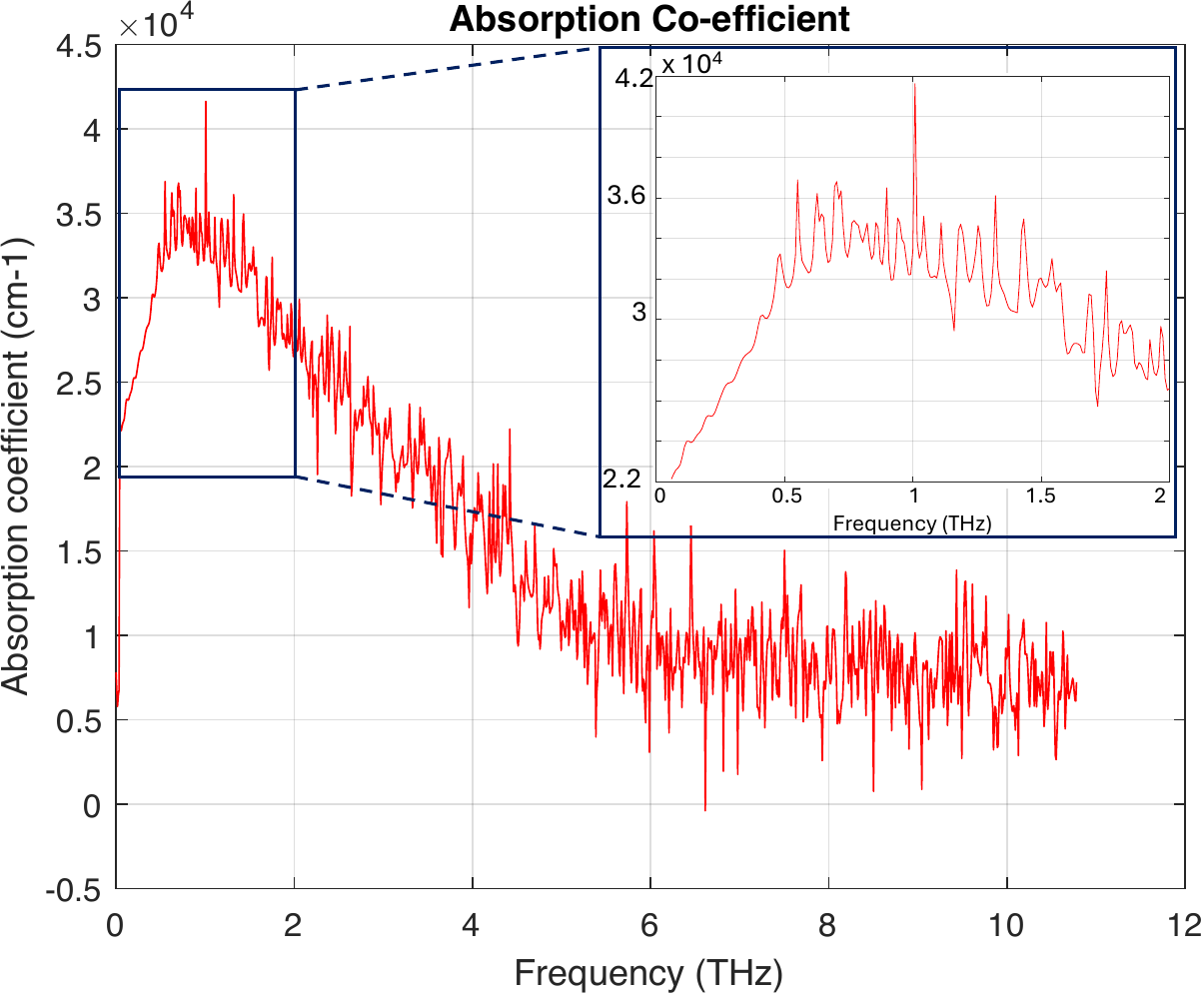}
        \caption{}
    \end{subfigure}
    \hfill
    \begin{subfigure}{0.47\linewidth}
    \centering
    \includegraphics[width=.93\linewidth]{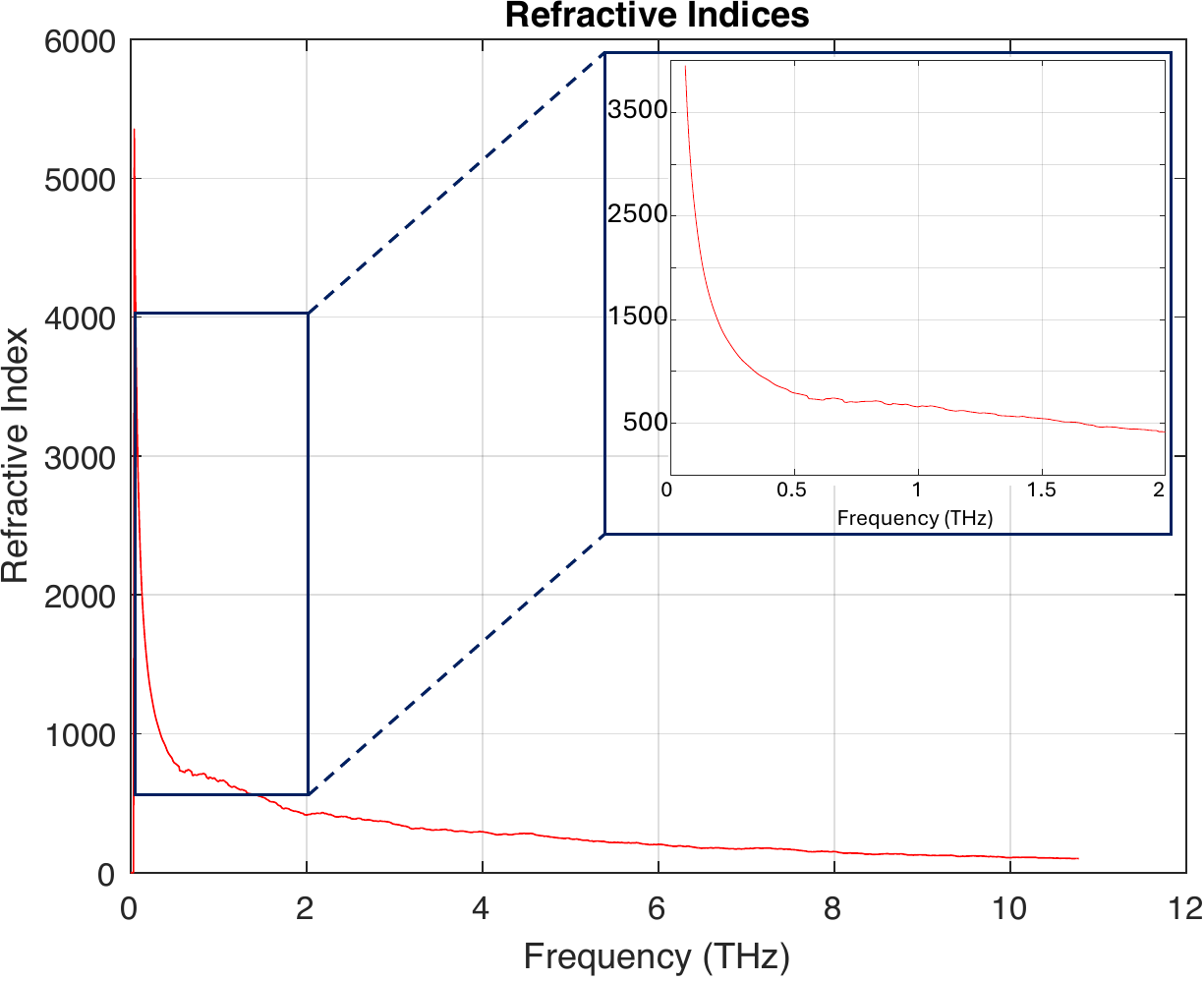}
    \caption{}
    \end{subfigure}
    \vspace*{-.3cm}
    \caption{a) Absorption coefficient of pig skin in the 0--12~THz range. b) refractive index of pig skin in the 0--12~THz range. A zoomed figure shows the behavior in a closed range around the frequency of the incident radiation.}
    \label{absorption}
\end{figure}

\begin{figure}[!b]
    \centering
    \begin{subfigure}{0.47\linewidth}
    \centering
    \includegraphics[width=.93\linewidth]{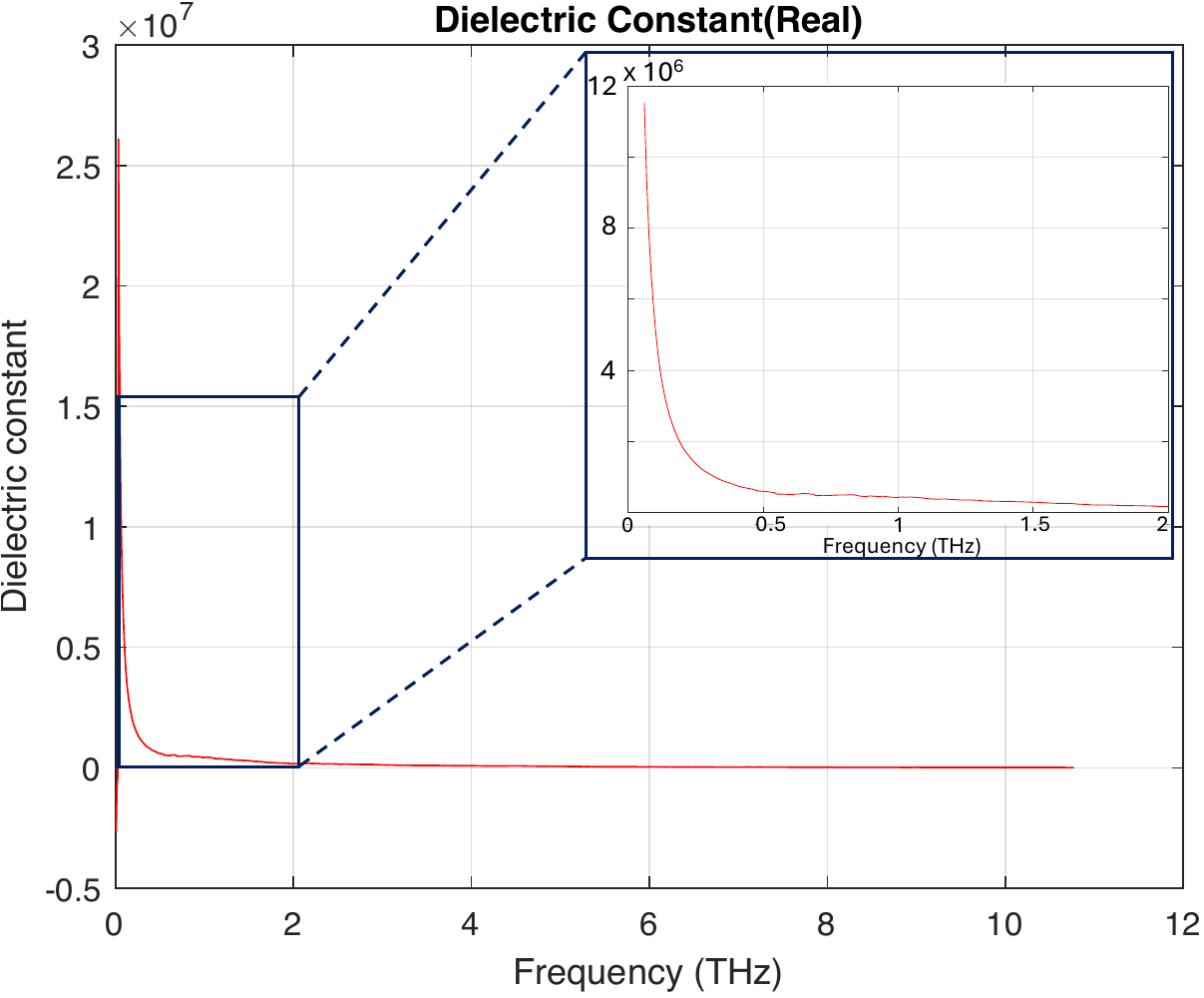}
    \caption{}
    \end{subfigure}
    \hfill
    \begin{subfigure}{0.47\linewidth}
        \centering
        \includegraphics[width=.93\linewidth]{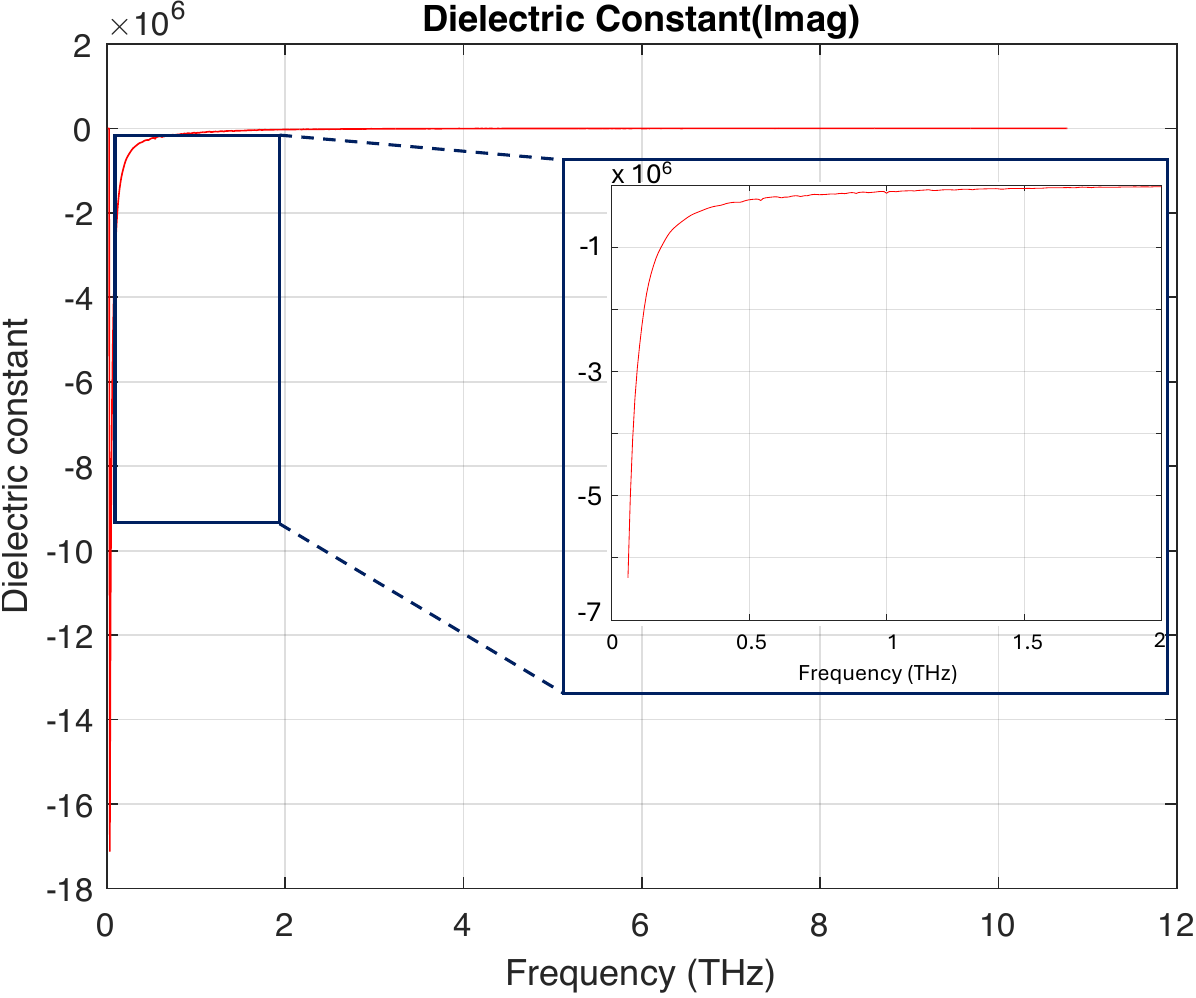}
        \caption{}
    \end{subfigure}
    \vspace*{-.3cm}
    \caption{Real (a) and imaginary (b) parts of the dielectric constant.
 A zoomed view shows the behavior near the incident radiation frequency.
}
    \label{die_cost}
\end{figure}

This section presents and discusses the numerical results for THz wave propagation through a layer of pig skin. The investigated frequency range extends from 0.01 to 11~THz, with the most relevant part zoomed to 2~THz in Figs.~\ref{trasm}--\ref{die_cost}, where the main dielectric relaxation processes occur. The electromagnetic response in this band is largely governed by tissue water content, affecting transmission, absorption, and dispersion.

Figure~\ref{trasm} presents the transmittance spectrum. Over most of the frequency band, transmission remains near zero due to strong absorption, with narrow peaks observed around 6–7~THz and smaller features between 7 and 9~THz. These narrowband spikes may originate from resonance effects such as Fabry–Pérot multi-reflections~\cite{Peretti2019} and from internal interference effects or from vibrational resonances of molecular constituents such as collagen and lipids~\cite{Nikitkina2021}. In fact, in this range, absorption is less dominated by dipolar relaxation of free water and more influenced by specific vibrational modes of bound water and structural proteins, leading to weak, narrowband variations in the transmittance spectrum. The inset up to 2~THz highlights the most physically relevant region, where transmittance rapidly decreases with increasing frequency as a result of the strong absorption of THz waves by the high water content of the tissue~\cite{jepsen2011terahertz,Arikawa2008,Saxena2024}.

From the transmittance spectrum, we extract the absorption coefficient $\hat{\alpha}(\omega)$, shown in Fig.~\ref{absorption}(a). The absorption first increases up to about 1~THz and then decreases with frequency. The strong low-frequency absorption (0–1~THz) reflects tissue hydration, as water molecules induce significant dielectric losses due to dipole relaxation~\cite{jepsen2011terahertz,Arikawa2008,Saxena2024}, confirming that hydration dominates the THz response of pig skin at low frequencies. The rapid oscillations may arise from intrinsic vibrational or rotational modes, numerical artifacts, or frequency-dependent scattering by tissue heterogeneities, such as adipose cells and collagen bundles~\cite{Nikitkina2021}. Beyond 2~THz, the absorption coefficient decreases further, indicating reduced dielectric losses as the response shifts from collective water relaxation to localized vibrational absorption. This trend aligns with previous studies on porcine and human skin, where absorption decays at higher THz frequencies, reflecting the transition from relaxation- to resonance-dominated behavior~\cite{Peralta2019,Sullivan2001,He2006,Helminiak2023,Piro2016,Nikitkina2021}.

Similarly, from transmittance we can also retrieve the refractive index $\hat{n}(\omega)$, shown in Fig.~\ref{absorption}(b). At very low frequencies, $\hat{n}$ reaches large values, then decreases steeply within the first 0.5~THz and gradually levels off at higher frequencies. When focusing on the 0.01--2~THz region, it is evident that the refractive index decreases monotonically with frequency: despite the abnormal high values due to internal normalization, the trend is consistent with previously reported results on water and biological tissues~\cite{Peralta2019}. Above 2~THz, the refractive index tends to stabilize toward lower values, approaching the asymptotic limit typical of bound water and macromolecular media~\cite{Nikitkina2021}. This plateau reflects the saturation of dispersive effects as the dipolar relaxation contribution vanishes, and the phase velocity becomes weakly frequency-dependent in the vibrational regime.

Overall, the results confirm that at low THz frequencies pig skin exhibits strong absorption and low transmittance, mainly due to its high water content. This behavior is consistent with the trends of $\hat{\alpha}(\omega)$ and $\hat{n}(\omega)$ for water predicted by the Debye model~\cite{Peralta2019}, highlighting the dominant role of hydration. From an application perspective, the results indicate that the upper part of the THz spectrum may be more favorable for imaging or sensing through biological tissues, as absorption decreases with frequency. Since THz radiation is strongly absorbed by liquid water, imaging applications can exploit the differences in water content between healthy and pathological tissues: for example, normal human skin typically contains about 70–72\% free water, while many cancers contain 82–85\%~\cite{Vilagosh2020}. From these trends, we can anticipate the behavior of the complex dielectric constant $\varepsilon(\omega) $$\,=\,$$ \varepsilon' - j \varepsilon''$. In particular, the results are consistent with a Debye-type dielectric dispersion, where molecular dipoles (primarily water) can follow the oscillating electric field at low frequencies, but fail to respond fast enough at higher THz frequencies~\cite{Peralta2019}.

As shown in Fig.~\ref{die_cost}, at low frequencies $\hat{\varepsilon}'$ reaches very large values, reflecting the strong tissue polarizability, while $\hat{\varepsilon}''$ is strongly negative, indicating significant absorption. As frequency increases toward the THz range (1--2~THz), $\hat{\varepsilon}'$ decreases and $\hat{\varepsilon}''$ rises toward zero, showing reduced absorption. Both $\hat{\varepsilon}'$ and $\hat{\varepsilon}''$ eventually stabilize at lower values. This trend aligns with the results for $\hat{\alpha}(\omega)$ and $\hat{n}(\omega)$, since $\hat{\varepsilon}''$ is directly related to the absorption coefficient and $\hat{\varepsilon}'$ to the refractive index. In the high-frequency region, both reach nearly constant values, indicating a steady-state dielectric response dominated by bound water and structural vibrations rather than collective dipole relaxation~\cite{Nikitkina2021}. This flattening of the permittivity spectra marks the transition from the Debye relaxation domain to the vibrational absorption domain above a few terahertz.

Furthermore, it is important to relate the dielectric response of porcine skin to the size of its structural components with respect to the THz wavelength. The main constituents, including keratinocytes, melanocytes, adipose cells, fibroblasts, and collagen fibers, have dimensions from a few to several tens of micrometers. Consequently, the observed spectral behavior primarily arises from the collective dielectric response of water and macromolecules rather than from scattering at cellular boundaries. This dimensional relationship explains why, in the 0 to 2~THz range, absorption and dispersion are dominated by Debye-type relaxation of free and bound water, while specific structural resonances become relevant only at higher frequencies. When the wavelength approaches the size of microstructures such as collagen bundles, local scattering and vibrational resonances can contribute to overall attenuation and dispersion~\cite{Nikitkina2021}.

In summary, the numerical results highlight the strong influence of water content on the THz response of pig skin, with high absorption and refractive index at low frequencies and reduced absorption at higher frequencies. The trends in $\hat{\alpha}(\omega)$, $\hat{n}(\omega)$, and $\hat{\varepsilon}(\omega)$ follow Debye-type behavior, confirming that hydration dominates the lower THz range. At higher frequencies, absorption and refractive index gradually decrease and stabilize, marking a transition to a less lossy yet dispersive regime. Despite low transmittance, frequency-dependent attenuation and phase delay provide a consistent description of wave propagation through hydrated skin. These parameters define key features of the THz channel, suggesting that selecting the operating frequency can optimize penetration and contrast, forming a basis for an analytical channel model capturing both dispersive and lossy characteristics relevant for intra-body communication and sensing.
\vspace*{-.3cm}

\section{Conclusions}\label{sec:conclusion}
This work presented an experimental characterization and modeling of porcine skin in the 0.01--11~THz range using PCA-based THz time-domain spectroscopy. The measured refractive index, absorption coefficient, and complex permittivity demonstrated that hydrated biological tissues behave as highly attenuating and dispersive media, with dielectric properties predominantly governed by water-related relaxation processes.

By mapping these frequency-dependent properties to a channel framework, a first experimental THz channel model was developed. The model captures the strong absorption at low frequencies, the transition to a less lossy dispersive regime at higher frequencies, and the associated phase delays, which are critical for accurate simulation of intra-body THz links and near-field sensing scenarios where local coupling dominates. These results extend previous studies by providing a quantitative, frequency-resolved description of the propagation channel, supporting the design and optimization of next-generation nanoscale biomedical devices operating in the sub-THz and THz bands.
\vspace*{-.3cm}

\subsubsection*{Acknowledgment}

The work of E. Marini, S. Mura, and M. Magarini was supported by the Institute of Information \& communications Technology Planning \& Evaluation (IITP) grant funded by the Korea government (MSIT)(No.RS-202400508465).

\bibliography{ref}
\bibliographystyle{IEEEtran}

\end{document}